\documentclass[aps,pra,showpacs,amsmath,amssymb,amsfonts,lengthcheck,twocolumn,longbibliography]{revtex4-1}

\usepackage{dsfont}
\usepackage{graphicx} 
\usepackage{graphics}

\usepackage{mathtools}
\usepackage{dcolumn}    
\usepackage{bm} 
\usepackage{graphicx}
\usepackage{amsmath}    
\usepackage{latexsym}
\usepackage{amsfonts}   
\usepackage{amssymb}
\usepackage{array}      
\usepackage{epsfig}
\usepackage{txfonts}
\usepackage{xcolor}
\usepackage[colorlinks=true,linkcolor=blue,urlcolor=blue,citecolor=blue,pdfusetitle]{hyperref}
\usepackage{hyperref}
\usepackage{ulem}

\newcommand {\fabs}[1] {\left| #1 \right|}
\newcommand{\norm}[1]{\left\lVert#1\right\rVert}

\newcommand{\ket}[1]{\left|#1\right\rangle}
\newcommand{\bra}[1]{\langle#1|}

\usepackage[none]{hyphenat}

\newcommand{\bla}{bla\\bla\\bla\\bla\\bla}
\begin{document}

\title{Minimal action control method in quantum critical models}
\author{Ainur Kazhybekova}
\author{Steve Campbell}
\author{Anthony Kiely}
\affiliation{School of Physics, University College Dublin, Belfield, Dublin 4, Ireland}
\affiliation{Centre for Quantum Engineering, Science, and Technology, University College Dublin, Belfield, Dublin 4, Ireland}
\begin{abstract}
We present a general protocol to control closed quantum systems that is based on minimising the adiabatic action. Using tools based on the geometry of quantum evolutions through the quantum adiabatic brachistochrone, we show that high fidelity control of the ground state of critical quantum systems can be achieved and requires only a reasonable approximation for the ground state spectral gap. We demonstrate our results for three widely applicable systems: the Landau-Zener, Ising, and fully connected spin models.
\end{abstract}
\date{\today}
\maketitle

\section{Introduction \label{intro}}
Achieving robust control of many-body quantum systems stands as an important step in order to fully realise quantum enhanced technologies \cite{DeutschPRXQ,OCreview2}. One natural approach involves exploiting adiabatic dynamics, where the system is manipulated sufficiently slowly that undesirable excitations are suppressed. While guaranteed to be effective, these techniques become highly impractical for many body systems where the requisite timescales will diverge as energy gaps close and therefore these methods fail once the evolution timescales are on the same order as the decoherence rate of the system.

Recently, significant steps have been taken in developing means to address this problem using the suite of tools from shortcuts to adiabaticity \cite{Torrontegui2013,Guery-Odelin2019} and optimal control theory \cite{OCreview,OCreview2}. The former, in particular, has been applied to several genuinely many-body systems and can achieve perfect control on arbitrarily short times \cite{delCampoPRL2012,Damski2014}. However, this comes at the price of highly non-local control fields which are impractical in terms of their experimental realisability and come with a diverging energetic cost \cite{Santos2015SciRep,Muga2017PRA,MugaNJP2018,Abah2019,LatunePRA,DeffnerEPL,CarolanPRA2022,Kiely2022}. In principle, the latter approach offers a more realistic route to control as experimental constraints can be built in {\it a priori}. Hybrid approaches can therefore offer remarkable versatility~\cite{Saberi2014,WhittyPRR,KielyNJP2021}.

For the ground state of a many-body system that is ramped through a quantum phase transition, the need for control can be traced back to the fact that as a system nears or crosses a critical point, spectral gaps shrink \cite{Caneva2011}. Different control methods circumvent this issue in distinct manners, for example counterdiabatic driving modifies the original energy spectrum such that vanishing energy gaps no longer occur and the time evolving state tracks the original Hamiltonian's instantaneous ground state, while in optimal control typically one is only concerned with achieving a particular target state and therefore avoids difficult regions of the parameter space by transiently leaving the adiabatic manifold. In this work we will explore an alternative to these approaches that is in the spirit of optimal control, requiring only local control fields, but that uses information about how the spectral gap behaves throughout the ramp protocol and therefore tracks an approximately adiabatic dynamics. Our protocol is derived from the quantum adiabatic brachistochrone by defining the adiabatic action~\cite{ActionPRL}, which takes into account the behaviour of the spectral gap. By minimising this action, an optimal ramp is determined and we demonstrate its effectiveness for three examples: first we recapitulate known results for the Landau-Zener model~\cite{CerfPRA}, then we apply these results and the minimal action technique to the Ising model where we demonstrate that the approach provides significant improvements over non-optimised ramps even if only an approximation to the spectral gap is used, and finally we examine a fully connected spin system.

\section{Minimal action approach to quantum control \label{prelim}}
%\subsection{Driving protocols}
We consider a quantum system described by the Hamiltonian $H(g)$, where $g$ is a scalar parameter which can be varied time-dependently. The adiabatic theorem provides the timescales over which the system can be driven while avoiding the generation of unwanted excitations and is inversely proportional to the minimal spectral gap. Therefore, when dealing with systems with vanishing energy gaps, as is characteristic of critical systems where a quantum phase transition occurs at $g\!=\!g_c$, the timescales necessary to achieve adiabatic dynamics tend to diverge. Thus, methods to manipulate such systems are desirable. This problem has particular relevance to the settings of quantum annealing ~\cite{brooke1999} and adiabatic quantum computation ~\cite{farhi2001}. Recently there has also been great focus on quantum heat engines which exploit criticality \cite{Cakmak2019,Fogarty2020}.

Here we take a heuristic approach by noticing that, since the minimal gap sets the timescale for achieving adiabatic dynamics, by judiciously varying the parameter $g$ taking into account the manner in which the spectral gap closes, high fidelity control can be achieved. This method is similar in spirit to several other approaches such as fast quasiadiabatic dynamics (FAQUAD) ~\cite{Martinez-Garaot2015,martinez2017,Torrontegui2019}, parallel adiabatic passage ~\cite{Guerin2011}, uniform adiabaticity~\cite{Quan2010} and local adiabatic evolution \cite{Richerme2013}. Note however that our method requires only energy eigenvalues, and not eigenvectors as in some other methods. This also relates to methods based on the relevant critical exponents of the model \cite{PolkovnikovPRL}.

Following the approach outlined in Ref.~\cite{ActionPRL}, we consider an adiabatic action defined as
\begin{eqnarray}
\label{action}
S=\int_0^\tau dt \frac{\hbar^2 ||\partial_t H(t) ||^2}{\Gamma^4(t)}.
\end{eqnarray}
where $\tau$ is the total duration of the protocol, $\Gamma(t)$ is the relevant energy gap, $|| \cdot ||$ is the Frobenius norm, and the choice of specific exponents is taken for convenience~\cite{ActionPRL}. The definition of $S$ is motivated by the usual adiabatic condition (which is often sufficient) \cite{Comparat2009}. Equally, $S$ provides a quantitative measure of the non-adiabaticity of a given protocol, having the units of time. Therefore, by minimising $S$ we can determine a ramp profile that keeps the system as close to adiabatic as possible for this action, i.e. by solving
\begin{equation}
\label{minaction}
    \frac{\delta S}{\delta g}=0
\end{equation}
for $g$. Note that the choice of $S$ ensures an Euler-Lagrange differential equation of second order which fits with the required boundary conditions: $g(0)=0$ and $g(\tau)=g_\tau$.

%We can view $S$ from a geometric perspective as defining a metric over the parameter manifold~\cite{}.

Equations~\eqref{action} and \eqref{minaction} form the basis of the control protocol. We will focus on the manipulation of the ground state, $\ket{\phi_0}$, for several systems and we will examine how a reasonable approximation for the ground state energy gap, $\Gamma$, is sufficient to achieve effective control, when compared with the performance of a simple linear ramp 
\begin{equation}
g(t)\!=\!g_0+ (g_\tau-g_0) t/\tau.
\end{equation}
The performance of each protocol will be quantified by focusing on the fidelity of the final state of the system, $\ket{\psi(\tau)}$, with the target ground state, i.e.
\begin{equation}
    \mathcal{F}=\fabs{\bra{\psi(\tau)}{\phi_0(\tau)}}^2.
\end{equation}
We assume $\ket{\psi(0)}\!=\!\ket{\phi_0(0)}$ as the initial condition in all cases i.e. the system starts in the ground state. To quantify the dynamics we use the instantaneous fidelity $\mathcal{F}_t=\fabs{\bra{\psi(t)}{\phi_0(t)}}^2$.

\section{Minimal action control: Case studies}

\subsection{Landau-Zener Model}
\label{LZM}
We begin with a pedagogical example, one which has been considered elsewhere in the literature~\cite{CerfPRA} using a similar framework, in order to outline the basic approach. Furthermore, due to its close relationship with the Ising model, the results of this section will be further employed when considering the many-body system in Sec.~\ref{TFIM}. Consider the Landau-Zener (LZ) model whose Hamiltonian is given by
\begin{equation}\label{eq:lzkz}
    H(t)=\hbar \Delta\sigma_x+\hbar g(t)\sigma_z,
\end{equation}
where $\Delta\!>\!0$ determines the minimal energy gap at the avoided crossing. While not a phase transition, this model nevertheless captures many of the essential features of a critical system, with spatial and dynamical critical exponents known to be $\nu=1$ and $z=1$, putting it within the same universality class as the Ising model. The energy gap between ground and excited states is given by
\begin{equation}
\label{LZgap}
\Gamma= \epsilon_1-\epsilon_0 = 2\hbar\sqrt{g(t)^2+\Delta^2}.
\end{equation}
The numerator of Eq.~\eqref{action} is readily determined as $\norm {\partial_t H}^2 \!=\! 2\hbar^2 \dot{g}^2$. Combining this with Eq.~\eqref{LZgap} we have that the
required action to be minimised, Eq.~\eqref{action}, is
\begin{eqnarray}
S=\int_0^\tau dt \frac{\dot{g}^2}{8(\Delta^2+g^2)^2}. \label{Slz}
\end{eqnarray}
The optimal ramp profile for $g(t)$ is then obtained by minimising the action Eq.~\eqref{minaction}, i.e. solving the Euler-Lagrange equation
\begin{equation}
\frac{\partial S}{\partial g} - \frac{d}{dt}\frac{\partial S}{\partial \dot g} = 0,
\end{equation}
which reduces to 
\begin{equation}
    \ddot{G}(s)=\frac{ 2G(s)\dot{G}(s)^2}{ \Delta^2 + G(s)^2},
    \label{eqn:LZM_EL}
\end{equation}
where we have made the substitution $s\!=\!\ t/\tau$ and defined $G(s)\!=\!g(s \tau)$. Equation~\eqref{eqn:LZM_EL} can be solved analytically. Fixing the initial and target conditions as $G(0)\!=\!g_0$ and $G(1)\!=\!-g_0$, i.e. symmetrically sweeping through the avoided crossing, we find the optimal ramp profile for $G(s)$ (equivalently $g(t)$) is
\begin{equation}
    G(s)=-\Delta \tan \left[(2 s-1)\arctan( g_0/ \Delta)\right].
    \label{eqn:LZM_gAction}
\end{equation}

\begin{figure*}[t]
\begin{center}
\includegraphics[width=0.3\textwidth]{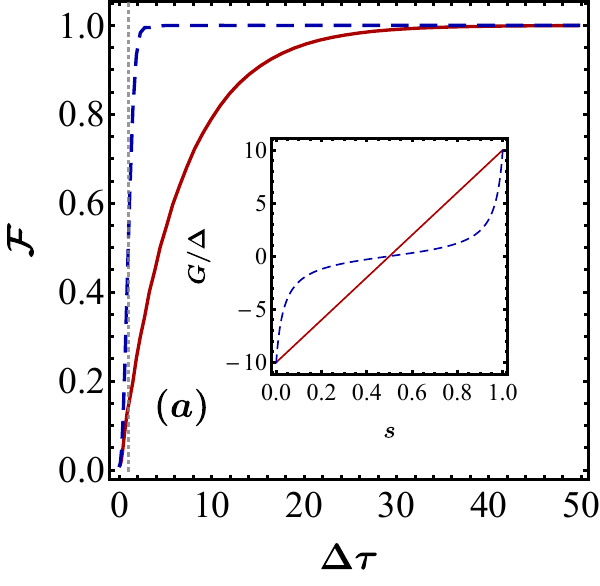} \qquad
\includegraphics[width=0.3\textwidth]{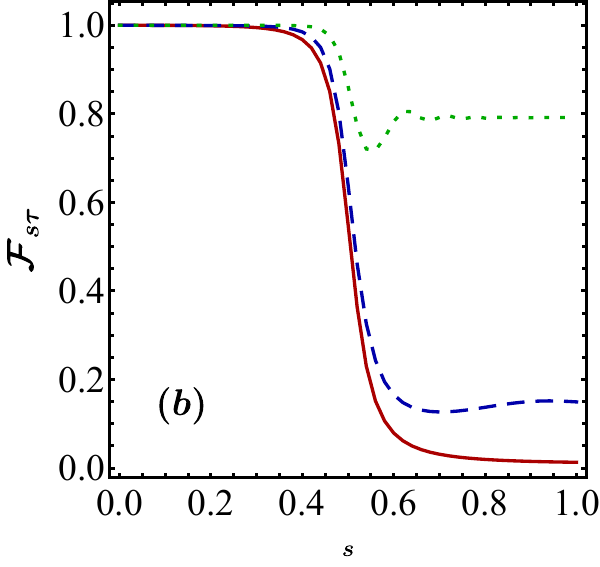}\qquad
\includegraphics[width=0.3\textwidth]{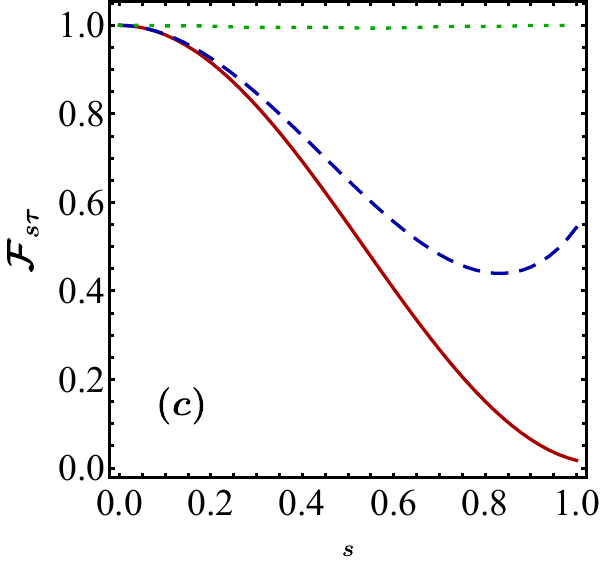}
\end{center}
\caption{Fidelity results for the Landau-Zener model. (a) The final fidelity $\mathcal{F}$ versus operation time $\tau$ for the linear ramp (red solid line) and minimal action method (blue dashed line). The light grey dotted line shows limiting timescale $\tau_a$. The inset shows the time variation of the two protocols. (b,c)  Instantaneous fidelity $\mathcal{F}_{s\tau}$ versus scaled time $s$ for the (b) linear ramp and (c) minimal action method. Total operation times are $\Delta\tau=0.1$ (red solid line), $\Delta\tau=1$ (blue dashed line) and $\Delta\tau=10$ (green dotted line). In all cases $g_0=-10 \Delta$.
\label{fig_LZ}}
\end{figure*}

% \begin{figure}[t]
% \begin{center}
% \includegraphics[width=0.95\columnwidth]{LZplot2.pdf}
% \end{center}
% \caption{ The final fidelity $\mathcal{F}$ versus operation time $\tau$ for the linear ramp (red solid line) and minimal action method (blue dashed line). \change{The light} \sout{Light} grey dotted line shows limiting timescale $\tau_a$ and $g_0=-10 \Delta$. The inset shows the time variation of the two protocols.
% \label{fig_LZ}}
% \end{figure}

% \begin{figure}[t]
% \begin{center}
% \includegraphics[width=0.49\columnwidth]{LZ_dynamicsA.pdf}\includegraphics[width=0.49\columnwidth]{LZ_dynamicsB.pdf}
% \end{center}
% \caption{\change{Instantaneous fidelity $\mathcal{F}_{s\tau}$ in the Landau-Zener model versus scaled time $s$ for the (a) linear ramp and (b) minimal action method. Total operation times are $\Delta\tau=0.1$ (red solid line), $\Delta\tau=1$ (blue dashed line) and $\Delta\tau=10$ (green dotted line).}\label{fig_lzydyn}}
% \end{figure}

% The energy eigenstates are
%  \begin{eqnarray}
%  \ket{\phi_0(t)} &=& \cos\left[\theta(t)\right]\ket{0}+\sin\left[\theta(t)\right] \ket{1} , \nonumber \\
%   \ket{\phi_1(t)} &=& \sin\left[\theta(t)\right] \ket{0}- \cos\left[\theta(t)\right]\ket{1}, \nonumber 
%  \end{eqnarray}
% where $\tan\left[\theta(t)\right]\!=\!-\left[g(t)+\sqrt{\Delta^2+g(t)^2}\right]/\Delta$ and 

In Fig. \ref{fig_LZ}(a) we show the performance of the optimised ramp, Eq.~\eqref{eqn:LZM_gAction}, where we see that by minimising the action the protocol achieves effectively adiabatic dynamics on timescales more than an order of magnitude faster than the linear ramp. Given that there is a finite minimum spectral gap, $\Delta$, no protocol can achieve a perfect fidelity in arbitrarily short times. As a rough estimate one cannot have perfect population transfer faster than the adiabatic timescale $\tau_a= \Delta^{-1}$. This is similar to results from quantum speed limits which give a timescale $\pi/(2 \Delta)$ ~\cite{HegerfeldtPRL,Caneva2009}. The inset shows the explicit profiles of the ramps and serves to elucidate why the action approach is so highly effective. Deviations from adiabaticity are dictated by the rate at which the spectral gap closes. In the case of the LZ model which hosts an avoided crossing at $g\!=\!0$, in order to maintain close to adiabatic conditions the ramp profile should necessarily reflect the behaviour of the spectral gap, $\Gamma$, in time. Thus, the action finds the ramp profile that most suitably tracks the behaviour of the spectral gap. For the LZ we see that this is characterised by rapid changes in the initial and final stages of the ramp, while as the system approaches the avoided crossing, $g(t)$ varies more gradually to account for the smaller spectral gap. 

In contrast, the linear ramp takes no account of how $\Gamma$ varies in time, and therefore insufficiently slow protocols lead to non-adiabatic dynamics as captured by the Kibble-Zurek mechanism~\cite{Kibble1976,Zurek1985,Damski2005}. In this specific case, the fidelity is well described by the Landau-Zener formula as $\mathcal{F}=1-\exp\left[- \pi \Delta^2 \tau/(2 |g_0|) \right]$.

To illustrate the difference in dynamics induced by the two protocols, we show the fidelity for both as a function of scaled time $s$ for different operation times $\tau$ in Fig.\ref{fig_LZ}. One can see clearly in Fig.\ref{fig_LZ} (b), that there is a sharp drop in instantaneous fidelity upon entering the impulse regime for the linear ramp \cite{CarolanPRA2022}. Nevertheless, for longer adiabatic timescales there is a revival after the dip near the avoided crossing \cite{MolmerPRA}. In contrast, in Fig.\ref{fig_LZ} (c) the results show that past the limiting timescale $\tau_a$ the minimal action method outperforms the linear ramp. In particular, for $\Delta \tau=10$ the state remains in the instantaneous ground state throughout the evolution.

% \begin{figure}[t]
% \begin{center}
% \includegraphics[width=0.95\columnwidth]{LZplot2.pdf}
% \end{center}
% \end{figure}

\subsection{Transverse-Field Ising Model}
\label{TFIM}

\begin{figure*}[t]
\begin{center}
\includegraphics[width=0.3\textwidth]{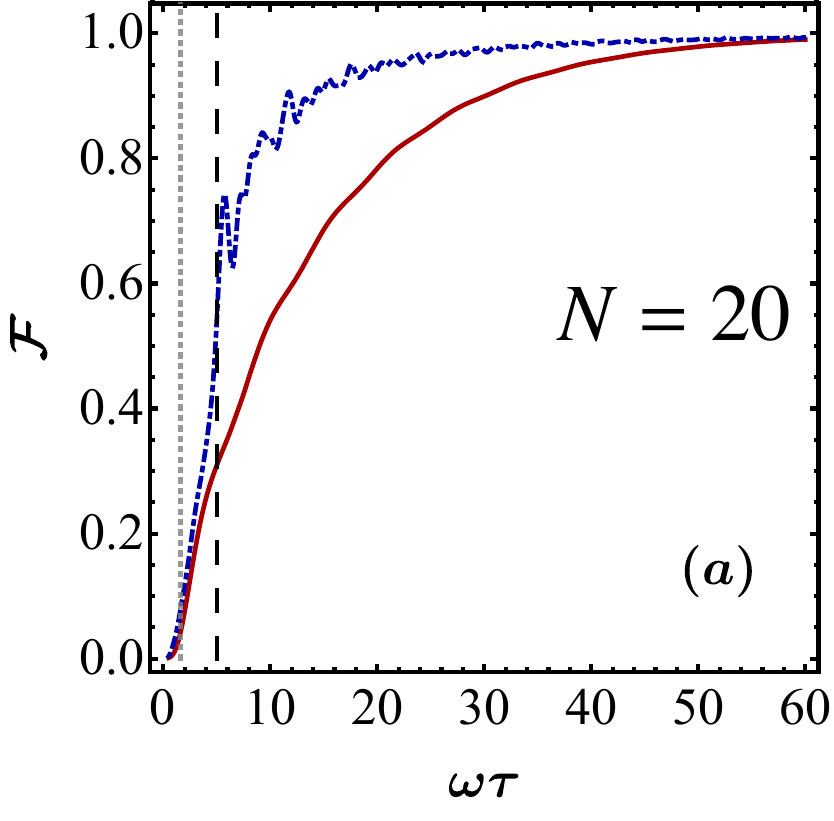}\qquad\includegraphics[width=0.3\textwidth]{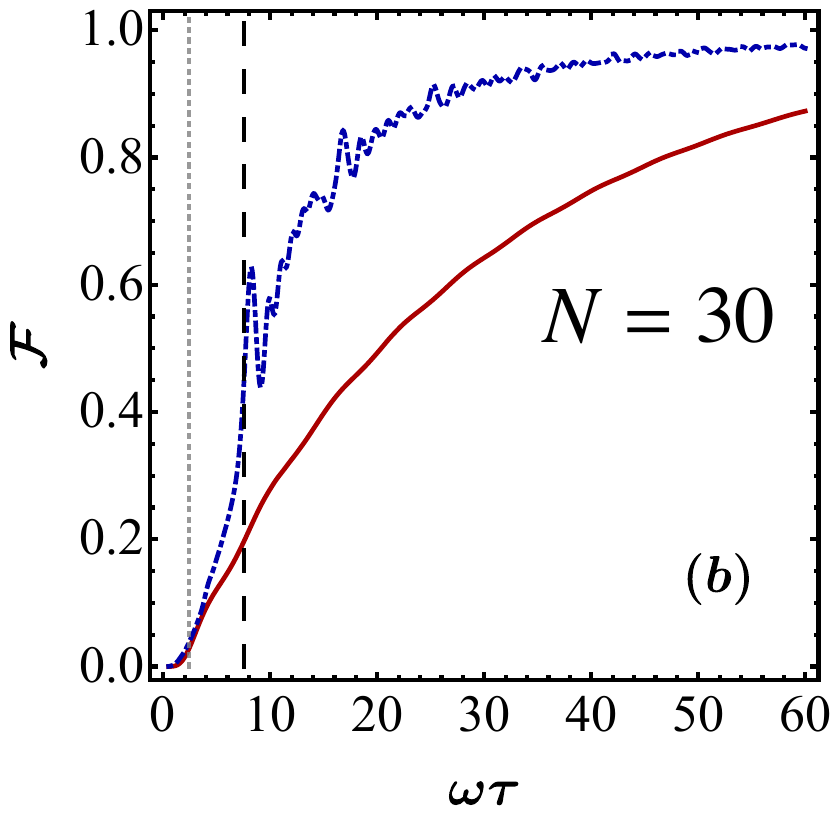}\qquad\includegraphics[width=0.3\textwidth]{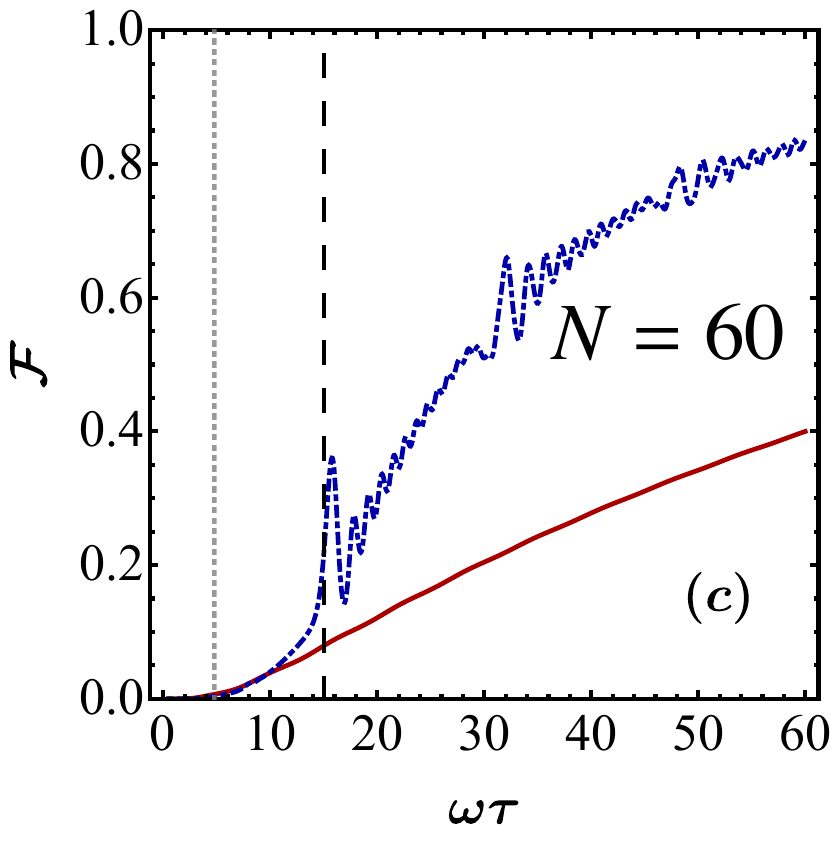}
\end{center}
\caption{ Final fidelity $\mathcal{F}$ versus operation time $\tau$ of the linear ramp (red solid line) and minimal action method (blue dashed line) for the Ising model with system size (a) $N=20$, (b) $N=30$, and (c) $N=60$. The approximations for the minimal timescales $\tau_a$ and $\tau_l$ are shown as light, dotted grey and dashed black vertical lines. We fix $g_0=0$. \label{fig_Ising}}
\end{figure*}

Moving now to a genuine many-body system, we consider the transverse field Ising model (TFIM) whose Hamiltonian is given by
\begin{equation}
    H(t)=- \hbar \omega\sum_{i=1}^{N}\left[g(t) \,\sigma_i^x+\sigma_i^z\sigma_{i+1}^z\right].
\end{equation}
and we assume periodic boundary conditions $\sigma_{N+1}^{x,y,z}\!=\!\sigma_1^{x,y,z}$ with an even number of sites $N$. Using the Jordan-Wigner transformation, we can map the model to $N/2$ non-interacting fermionic momentum subspaces labelled by the discrete momentum values $k=(2n-1)\pi/(N b)$ where $b$ is the distance between spins and $n=1, \ldots, N/2$. Each momentum subspace is an effective Landau-Zener model with a relevant spectral gap given by $\Gamma_k\!=\!4 \hbar \omega \sqrt{g(t)^2-2g(t)\cos(kb)+1}$ which vanishes in the thermodynamic limit at the critical point $g_c=1$ \cite{Puebla2020}. Thus, while the model is in principle exactly solvable, our interest will be in demonstrating that the minimal action approach can still be highly effective even if details of the exact spectral gap are not known~\cite{ActionPRL}. 

%\sout{In general this action can be written as}
Thus, instead of looking at the action for the exact ground state spectral gap, we consider the adiabatic action for only the lowest momentum subspace ($n=1$), which is the most relevant in terms of the model's critical features. The general action for a system of this form can be written as
\begin{eqnarray}
\label{actionTFIM}
S=\int_0^\tau dt \left[ \frac{\alpha \dot{g} \gamma^2}{(g-\beta)^2+\gamma^2} \right]^2.
\end{eqnarray}
The corresponding Euler-Lagrange equation in this case is
\begin{eqnarray}
\ddot{G}=\frac{2 (G-\beta ) \dot{G}^2}{\beta ^2+\gamma^2-2 \beta  G+G^2}.
\end{eqnarray}
This gives the optimal solution
\begin{eqnarray}
\label{SolnTFIM}
G(s)=\beta+\gamma \tan \left[(1-s) \tan^{-1}\left(\frac{g_0-\beta}{\gamma}\right)-s \tan^{-1}\left(\frac{g_0+\beta-2}{\gamma}\right) \right], \nonumber \label{opt} \\
\end{eqnarray}
where $\beta$ is related to the point of avoided crossing (similar to $g_c$) and $\gamma$ is related to the minimal energy gap of the subspace.

Due to the decoupling into individual momentum subspaces, we see that the action, Eq.~\eqref{actionTFIM}, has the same form as the Landau-Zener model action, Eq.~\eqref{Slz}. This can be seen explicitly by identifying $\alpha=1/(2 \sqrt{2}\Delta^2)$, $\beta=g_c=0$ and $\gamma=\Delta$ for this case.

However, for the Ising model the location of the critical point is different. Thus we can extract the action solution by identifying that $\alpha\!=\!\left[4 \sqrt{2}\omega \sin^2(\pi/N)\right]^{-1}$, $\beta=\cos(\pi/N)$ and $\gamma=\sin(\pi/N)$. Using these in Eq.~\eqref{actionTFIM}, we find the specific action for the Ising model is given by
\begin{eqnarray}
S=\int_0^\tau dt \frac{  \dot{g}^2}{32 \omega^2 [g^2-2g \cos(\pi/N)+1]^2}.
\end{eqnarray}

We demonstrate the effectiveness of this approach in Fig.~\ref{fig_Ising} for several different system sizes compared to a simple linear ramp. As expected, larger system sizes necessarily imply that longer time scales are required to approach the adiabatic limit. Regardless of system size, we clearly see that the linear ramp fails to achieve the target state, with its performance rapidly dropping off even for modest increases in system size, cfr. panels (a) and (b). By using the minimal action approach applied to only the lowest momentum subspace, we see that significantly higher fidelities can be achieved on fast timescales, although remark that there is still a fundamental lower limit to how fast the protocol can be achieved. An approximation for this minimal timescale is given by adiabatic timescale $\tau_a= 1/\left[4\omega \sin(\pi/N)\right]$ which varies extensively for large $N$. Note that for the linear protocol, the scaling of the minimum time for adiabatic evolution is quadratic in $N$ ~\cite{DziarmagaPRL}. While in the case of the LZ model we find the minimal action protocol almost saturates this minimal time, we note there is a larger discrepancy here in the case of the Ising chain. This is due to locality effects. From the periodic boundary conditions, the largest distance between spins is $N b/2$, while the maximal group velocity is $2 \omega b$ ~\cite{Najafi2017}. This gives a heuristic timescale $\tau_l\!=\! N/(4 \omega)$, shown in Fig. \ref{fig_Ising} as a vertical black dashed line.  Note that there are sharp peaks of high fidelity near $\tau=\tau_l$.

% From the periodic boundary conditions, the largest distance between spins is $N b/2$, while the maximal group velocity is $2 g \omega b$ for $g<1$ and $2 \omega b$ for $g>1$ ~\cite{}. This gives two timescales $N/(4 \omega g)$ and $N/(4 \omega)$ (is there a way to average these??). In Fig. \ref{fig_Ising} we include the \steve{best case<-what does that mean?} scenario timescale $\tau_l=N/(4 \omega)$ as a vertical dashed line.  Note that there are sharp peaks of high fidelity near $\tau=\tau_l$.

% For all sizes we see that $\tau_l$ provides a useful timescale for when the first significant increase in final state fidelity is achieved.The performance of the minimal action protocol demonstrates the relevance of $\tau_l$ where at \steve{multiples?? what's the relation} the protocol obtains `sweet spots' in ramp duration where higher fidelities can be achieved. 

By only optimising over the lowest momentum subspace, we see that while not perfectly effective, significant gains in the target state fidelity can still be achieved. Thus, the application to the Ising model serves to demonstrate that one does not necessarily require precise knowledge of the spectral gap, i.e. $\Gamma$ entering Eq.~\eqref{action}, it is already sufficient to have a reasonable estimate in order to find protocols that can perform significantly better.

\begin{figure}[t]
\begin{center}
\includegraphics[width=0.49\columnwidth]{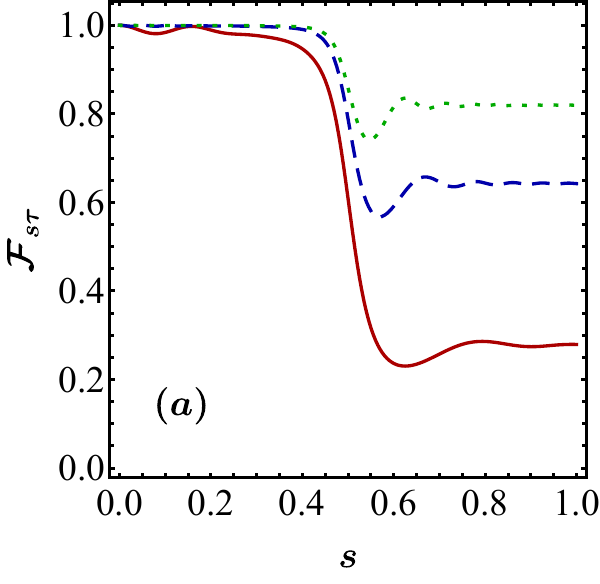}\includegraphics[width=0.49\columnwidth]{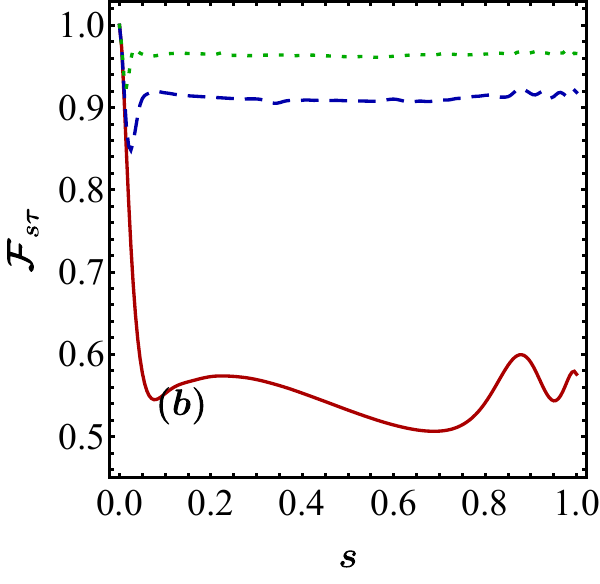}
\end{center}
\caption{Instantaneous fidelity $\mathcal{F}_{s\tau}$ in the Ising model versus scaled time $s$ for the (a) linear ramp and (b) minimal action method. Total operation times are $\omega\tau=10$ (red solid line), $\omega\tau=30$ (blue dashed line) and $\omega\tau=50$ (green dotted line), the system size is $N=30$ and $g_0=0$.\label{fig_isgdyn}}
\end{figure}

To understand the dynamical evolution of the state for each protocol, we consider the intantaneous fidelity in Fig.\ref{fig_isgdyn}. For the linear protocol, in Fig.\ref{fig_isgdyn}(a), we once again see a steep drop in fidelity crossing the critical point. In contrast, for the minimal action protocol (shown in Fig. \ref{fig_isgdyn}(b)), there is an immediate dip in fidelity. This is related to excitations in the higher momentum subspaces which have avoided crossings for values less than $g_c$. The minimal action protocol traverses these avoided crossings faster than the linear ramp, thus inducing excitations in these subspaces leading to this behaviour. This highlights the potential benefits of including the next highest momentum subspace.

\subsection{Fully connected model}
\begin{figure*}[t]
\begin{center}
\includegraphics[width=0.31\textwidth]{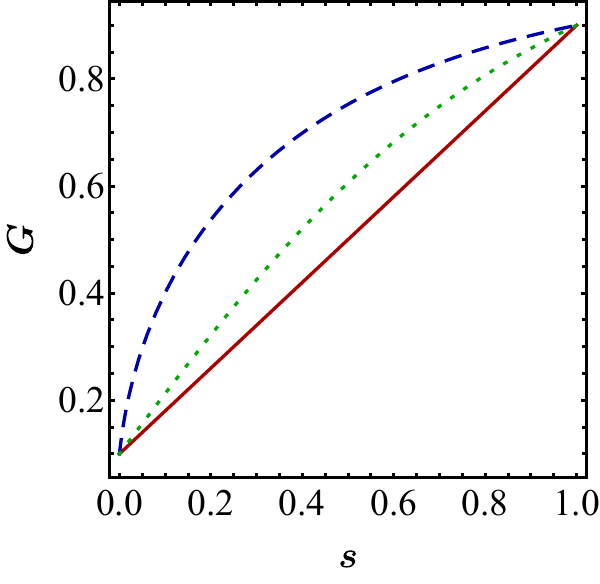}\includegraphics[width=0.31\textwidth]{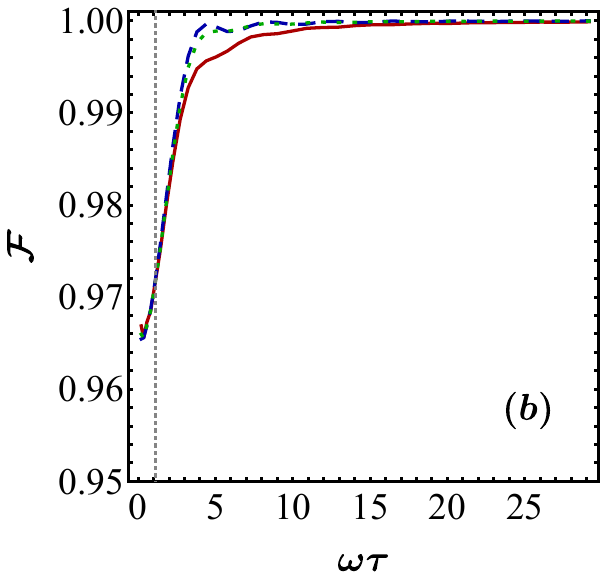}~~\includegraphics[width=0.31\textwidth]{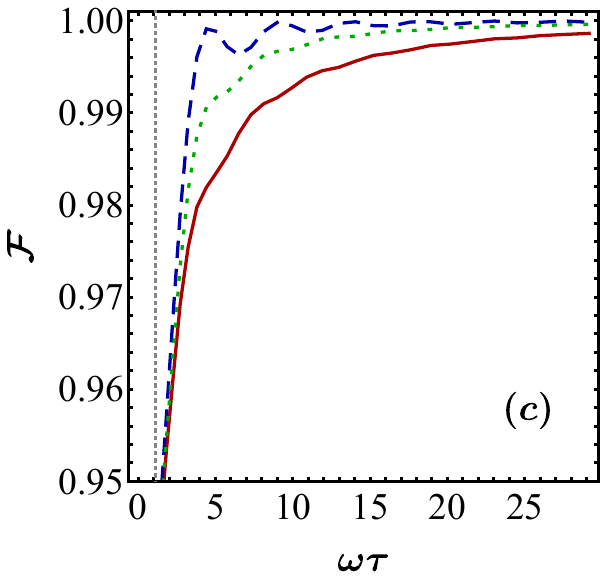}
\end{center}
\caption{Comparison of protocols: linear ramp (red solid line), minimal action method (blue dashed line) and alternative scheme (green dotted line). Panel (a) shows the temporal variation of the different schemes. The resulting final fidelities versus operation time is shown for (b) $\eta=10$ (c) $\eta=100$. Timescale $\tau_a$ shown as light grey vertical dotted line with parameters  $g_0=0.1$ and $g_\tau=0.9$\label{fig_connected}.}
\end{figure*}
As a final model, we move beyond the universality class captured by the Ising model and consider a fully connected spin system whose Hamiltonian is given by~\cite{Garbe2022}
\begin{equation}
H(t) = \hbar \omega a^{\dagger} a - \frac{\hbar \omega g^2(t)}{4} \left(a + a^{\dagger} \right)^2 + \frac{\hbar \omega g^4(t)}{16 \eta} \left(a + a^{\dagger} \right)^4,
\end{equation}
with bosonic creation and annihilation operators $a^{\dagger}$ and $a$, restricted to the regime $0 < g(t) < 1$. This model is widely applicable as a low-energy effective Hamiltonian for systems with infinite range interactions such as the Quantum Rabi model~\cite{RicardoPRL} or Lipkin-Meshkov-Glick model~\cite{Garbe2022} where $\eta$ is an effective system size. 

This model has a critical point at $g_c\!=\!1$ in the thermodynamic limit $\eta\! \rightarrow\! \infty$ with critical exponents given by $\nu z \!=\! 1/2$. In the thermodynamic limit the relevant energy gap is given by $\Gamma=2 \hbar \omega \sqrt{1-g^2}$ \cite{Garbe2022}. For sufficiently large systems, this serves as a reasonable estimation of the true spectral gap. Using this in Eq.~\eqref{action} and neglecting irrelevant constant factors, the action is given by
\begin{eqnarray}
S = \int_0^\tau dt \frac{g^2 \dot{g}^2}{\omega^2(1-g^2)^2}.
\end{eqnarray}
For this case the Euler-Lagrange equation becomes
\begin{eqnarray}
\ddot{G}(s) = \left [ \frac{G(s)^2+1}{G(s)^2-1} \right] \frac{\dot{G}(s)^2}{G(s)}.
\end{eqnarray}
Solving this with the appropriate boundary conditions we find the optimised profile
\begin{eqnarray}
G(s)= \sqrt{\left(g_0^2-1\right)
   \left(\frac{g_\tau^2-1}{g_0^2-1}\right)^s+1},
\end{eqnarray}
shown in Fig.\ref{fig_connected}(a). For comparison, we also consider another ramp profile determined in Ref.~\cite{Garbe2022} based on the critical exponents
\begin{eqnarray}
G(s)=\sqrt{2-\frac{2}{s^2+1}} (g_\tau-g_0)+g_0.
\end{eqnarray}

The resulting fidelities are compared with the linear case in Fig.\ref{fig_connected} (panels (b) and (c))  for different values of the effective system size $\eta$. The adiabatic timescale  $\tau_a=1/\left[2 \sqrt{1-g_{\rm max}}\right]$ marks a precipitous drop in fidelity where $g_{\rm max}$ is the maximal value of $g$ during the evolution. In our setting $g_0 < g_\tau$ so $g_{\rm max}=g_\tau$. Note that there are no locality effects in this setting since every spin interacts directly with every other spin.

For small system sizes ($\eta=10$), the minimal action protocol provides only modest improvements over the linear ramp. However for increasing system size, the approximation to the spectral gap becomes more accurate. Hence for larger system sizes ($\eta=100$) we see that the minimal action protocol significantly outperforms the other two protocols.

% \change{We also show the dynamics in this case [c.f. Fig. \ref{fig_hcondyn}] which differs significantly from the previous two examples due to differing critical exponents and the fact that the critical point is only approached in this case. This is apparent in Fig. \ref{fig_hcondyn}, which does not display the characteristic dip as in previous examples. Note also the the minimal action method [c.f. Fig. \ref{fig_hcondyn}(b)] smoothly approaches the target state (and hence is more robust to timing errors) with higher fidelity.}

% \begin{figure}[t]
% \begin{center}
% \includegraphics[width=0.49\columnwidth]{Hcon_dynamicsA.pdf}\includegraphics[width=0.49\columnwidth]{Hcon_dynamicsB.pdf}
% \end{center}
% \caption{\change{Instantaneous fidelity $\mathcal{F}_{s\tau}$ in the fully connected model versus scaled time $s$ for the (a) linear ramp and (b) minimal action method. Total operation times are $\omega\tau=2$ (red solid line), $\omega\tau=5$ (blue dashed line) and $\omega\tau=10$ (green dotted line) the effective system size is $\eta=100$, $g_0=0.1$ and $g_\tau=0.9$.}\label{fig_hcondyn}}
% \end{figure}

\section{Conclusion \label{con}}
We have provided a general protocol for finding (near) optimal ramps to control critical quantum systems. The scheme is based on the adiabatic action, Eq.~\eqref{action} and is applicable to any model where a good approximation of the relevant spectral gap is known. We demonstrated the utility of the minimal action control methods with three paradigmatic examples. For the Landau-Zener model, the optimal solution is readily obtainable~\cite{Tomka2016, CerfPRA} and served to demonstrate the underlying principle that in such models, high fidelity local control is obtainable provided the ramp profiles suitably take into account information regarding the spectral gap. The versatility of the approach was then demonstrated for the Ising model, where rather than the precise ground state spectral gap, an approximation using only the lowest momentum subspace was used to define the adiabatic action. Significant improvements in target state fidelity were achieved, and thus this serves to demonstrate that only a reasonable estimate, and not the precise manner, regarding how the energy gaps close is needed. For the Ising model, we expect that the results could be improved by including the effects of the higher momentum subspaces. For the $n^{\rm th}$ subspace, the optimal solution is given by Eq. \eqref{opt} with generalised parameters $\alpha_n \propto 1/\gamma_n^2$, $\beta_n=\cos\left((2n-1)\pi/N \right)$ and $\gamma_n=\sin\left((2n-1)\pi/N \right)$. These could be combined for instance via a weighted average of the minimal gaps $\gamma_n^{-1}$. Finally, we showed that the minimal action approach applies equally well for critical systems in other universality classes by considering a fully connected spin system, where high target state fidelities were achieved based on an approximation for the relevant energy gap. This broadens the approach’s applicability to new models where the gap can only found numerically, but with a scaling that can be modelled by a simple analytical function.

Our results add to the growing body of work highlighting that a hybrid control techniques exploiting both analytic and numeric aspects provide a viable route to achieve coherent control. In particular, our results may find fertile application in more complex models provided an estimation of the relevant spectral properties can be determined. This has relvance to the development of approximate counterdiabatic driving protocols for many-body systems \cite{Ieva2022}.

Furthermore, they may also prove relevant for a more diverse range of areas. These include open systems~\cite{Santos2021}, the energetic cost of quantum gates operations~\cite{DeffnerPRE2017}, generalisations of the adiabatic theorem~\cite{KosloffPRR2021}, quantum refrigerators using working mediums which posses an uncontrollable finite gap~\cite{KosloffPRE2010}, thermodynamic protocols with minimal dissipation~\cite{CrooksPRL} and quantum speed limits where action based formalism has been previously developed~\cite{OConnorPRA2021}.

%\sout{These include open systems~\cite{Santos2021}, quantum thermodynamics~\cite{DeffnerPRE2017} (where protocols with minimal dissipation are desired~\cite{CrooksPRL}) and quantum speed limits where action based formalism has been previously developed~\cite{OConnorPRA2021}.}

% Note that this approach can be generalised to the case of open systems \cite{Santos2021}.

% These ideas can also prove relevant in the case of thermodynamics~\cite{DeffnerPRE2017}, for example for protocols which minimise dissipation ~\cite{CrooksPRL}.

% Quantum speed limits have previously been developed using an action based formalism \cite{OConnorPRA2021}. The presented method 

\acknowledgements
This work is supported by the Science Foundation Ireland Starting Investigator Research Grant ``SpeedDemon" No. 18/SIRG/5508 and the John Templeton Foundation (Grant ID 62422).

% \change{
% \appendix

% \section{Higher momentum subspaces of Ising model \label{appA}}

% }

\bibliography{control.bib}
\end{document}